# Transient Negative Capacitance in Ferroelectric Nematic Liquid Crystals

Netra Prasad Dhakal[1,2], Alex Adaka[1,2], Robert J. Twieg[4], Antal Jákli[1,2,3]*

[1]Advanced Materials and Liquid Crystal Institute, Kent State University, Kent, OH 44242, USA

[2]Materials Science Graduate Program, Kent State University, Kent, OH 44242, USA

[3]Department of Physics, Kent State University, Kent, OH 44242, USA

[4]Department of Chemistry and Biochemistry, Kent State University, Kent, OH 44242, USA

*: Corresponding author: ajakli@kent.edu

**Abstract:**

The transient negative capacitance (NC) of solid ferroelectric materials used in field effect transistors can reduce the power dissipation of electronics. Here we show that similar negative capacitance appears in the recently discovered fluid ferroelectric nematic liquid crystal (FNLC) films while switching their ferroelectric polarization. Instead of sidewise motion of domains with opposite direction of polarization of ferroelectric crystals, in FNLCs the polarization vector that at zero field aligns parallel to the substrates, rotates toward the applied electric field. In spite of that difference, the observed switching time per unit electric field values are comparable. Although current FNLCs have much larger electric conductivity than of ferroelectric crystals, they have several advantages: they are organic, lightweight, can be prepared and operated at ambient temperature, the direction of the polarization can be set by simple surface alignment, and they require much less electric field for switching of the polarization. These properties make FNLCs attractive to test them in in flexible organic electrochemical transistor (OECT) devices.

# I. Introduction

A capacitance is usually positive, i.e., an increasing potential difference leads to an increasing charge stored on the electrodes. Negative differential capacitance of ferroelectric films has been predicted already the early days of ferroelectricity [1]. Experimentally a transient negative slope of the polarization-electric field dependence, suggesting a transient negative capacitance (NC) in thin BaTiO$_3$ ferroelectric capacitors was observed only in 2006.[2]

The principle of NC can be understood by starting out from the definition of the capacitance $C = dQ/dV$ where $Q$ is the charge stored on the electrodes and $V$ is voltage between them. A negative capacitance means a decrease of the potential difference ($dV < 0$) while the charge stored on the electrodes increases ($dQ > 0$).[3,4] For a dielectric material the electric displacement $\vec{D}$ is proportional to the applied field $\vec{E}$ as $\vec{D} = \varepsilon_o \varepsilon_r \vec{E}$, where $\varepsilon_o = 8.85 \cdot 10^{-12} \frac{C^2}{Nm^2}$ is the permittivity of vacuum, and $\varepsilon_r$ is the relative dielectric permittivity of the material, which is always positive. A ferroelectric material has a spontaneous polarization $\vec{P}_s$ and $\vec{D} = \varepsilon_o \varepsilon_r \vec{E}_F + \vec{P}_s$, where $\vec{E}_F = \vec{E} - \frac{\vec{P}_{s\prime}}{\varepsilon_o \varepsilon_r}$ is the electric field in the ferroelectric material. This provides that

$$\frac{\partial \vec{D}}{\partial \vec{E}_f} = \varepsilon_o \varepsilon_r + \frac{\partial \vec{P}_s}{\partial \vec{E}_f} \approx \frac{\partial \vec{P}_s}{\partial \vec{E}_f} \equiv \varepsilon_o \varepsilon_F, \qquad (1)$$

where $\varepsilon_F$ is the ferroelectric permittivity. According to Landau model, the free energy density of a ferroelectric material with a magnitude of spontaneous polarization $P_s$ in an electric field $E_F = |\vec{E}_F|$ can be written as

$$f = aP_s^2 + bP_s^4 - E_F P_s \qquad (2)$$

Here $a < 0$ and $b > 0$ are the Landau coefficients assuming second order paraelectric – ferroelectric phase transition. For $E_F = 0$ this corresponds to a double-well function with stable polarization values at the minima and an unstable $P_s = 0$, as depicted in Figure 1(a). Minimizing $f$ with respect to $P_s$, we get for the electric field in the ferroelectric materials that

$$E_F = 2aP_s + 4bP_s^3, \qquad (3)$$

which has an "S" shape as shown in Figure 1(b). In this curve the orange shaded thermodynamically unstable range represents $\varepsilon_F < 0$ where the polarization changes opposite to the electric field, therefore the capacitance becomes negative. [3] If we apply an external square shape voltage $V_S$ in a circuit consisting of a resistance $R$ connected in series with a ferroelectric material represented by a polar capacitance and measure the voltage drop on the ferroelectric material $V_F$ (see Figure 1(c)), we expect to find a dip in the time dependence of $V_F$ (red curve as shown in Figure 1(d)).[5]

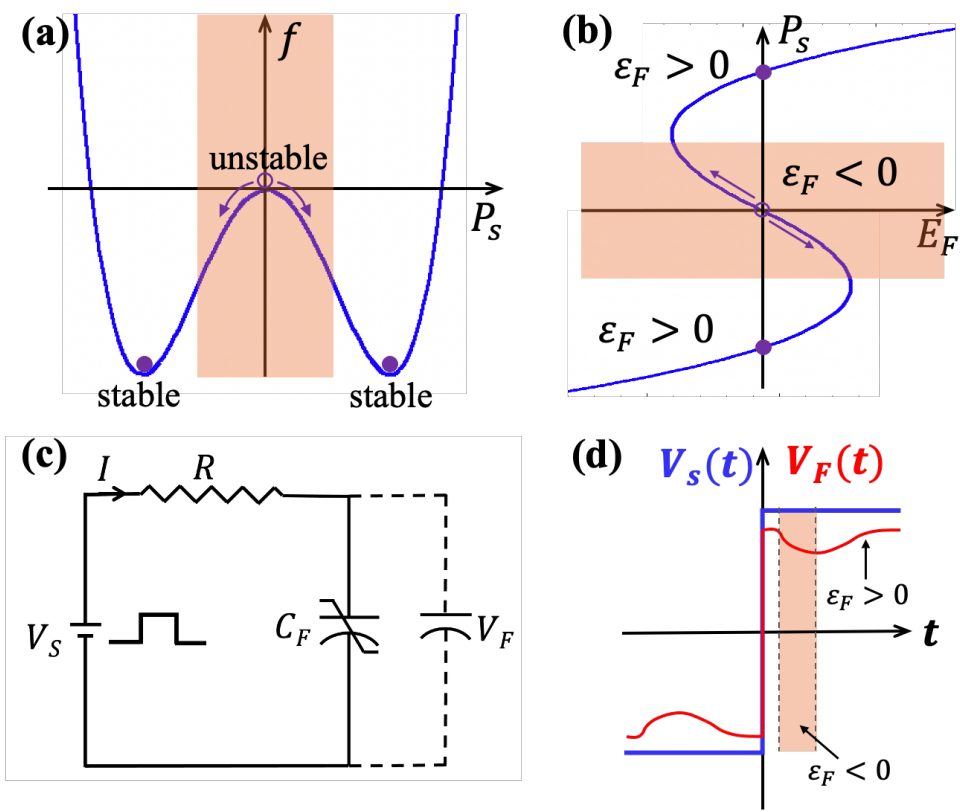

Figure 1:(a): Schematic free energy density $f$ as a function of the ferroelectric polarization $P_s$ according to Landau theory of a ferroelectric materials with second order paraelectric-ferroelectric phase transition. (b): Ferroelectric polarization $P_s$ as a function of the electric field in a ferroelectric material with $f(P_s)$ corresponding to (a). (c): Schematic representation of a circuit consisting of an external voltage source $V_S$, a resistance $R$ connected in series with an insulating ferroelectric material represented by a polar capacitance, and a voltmeter that measures the voltage drop on the ferroelectric material $V_F$. (d): Schematic time dependence of the voltage $V_F$ (red curve) flowing through a ferroelectric material under stepwise $V_s$ voltage applied to the ferroelectric film. Orange shaded areas indicate unstable state.

NC of ferroelectric materials used in field effect transistors can reduce the power dissipation of electronics[6], although most of ferroelectric solids cannot be integrated with current semiconductor fabrication technologies[3].

The first direct measurement of transient NC was reported in 2016 in gadolinium (Gd) doped ferroelectric hafnium oxide ($HfO_2$) thin film in metal-ferroelectric-metal capacitor and a resistor arrangement[7]. Later a number of other solid ferroelectrics were found to show similar behavior. [5,8–10] These results clearly show that NC-FET ferroelectrics will pay an important role in future electronic materials.

Very recently it was shown that the NC effect in thin films of trimethyl chloromethyl ammonium trichloro cadmium (II) molecular ferroelectric capacitance devices was comparable to that of conventional $HfO_2$-based ferroelectrics.[11] This is significant, since molecular ferroelectrics can be flexible, biocompatible, lightweight, require low preparation temperature and their properties can be tuned by chemical design, thus making promising alternatives for future electronic devices.

In this paper we extend the study of transient NC effect to fluid molecularly ferroelectric materials, namely the recently discovered ferroelectric nematic liquid crystals (FNLCs) [12–18]. They have promising electro-optical properties[19,20], they form fluid filaments[21,22], possess piezoelectricity[23], and in their chiral form they can also be used as electrically tunable reflectors [24–26] and lenses.[27] FNLCs can also form relaxor ferroelectrics when doped in non-polar materials.[28] They are lightweight, can be prepared and operated at ambient temperature and the direction of the polarization can be set by simple surface alignment.[29]

Nematic liquid crystal (NLC) is a state of matter used in today's omnipresent flat panel LCD displays, having a long-range orientational order without a positional order. The average orientation of molecules is apolar and denoted by the director ($\hat{n} = -\hat{n}$) (Figure 2(a)).

Ferroelectric Nematic Liquid Crystals (FNLCs) have strongly polar (dipole moments $\geq 9\ D$)[18,30,31] molecules whose long axis and dipole moments align macroscopically the same direction, resulting in a polar director ($\hat{n} \neq -\hat{n}$) and macroscopic dipole density $\vec{P}_S = P_S \cdot \vec{n}$ (Figure 2(b)). Instead of sidewise motion of domains with opposite direction of polarization of ferroelectric crystals, in FNLCs the polarization vector that at zero field aligns parallel to the substrates, rotate toward the electric field applied, competing with the depolarization field $E_d = -\frac{P_S}{\varepsilon_0 \varepsilon_F}$.

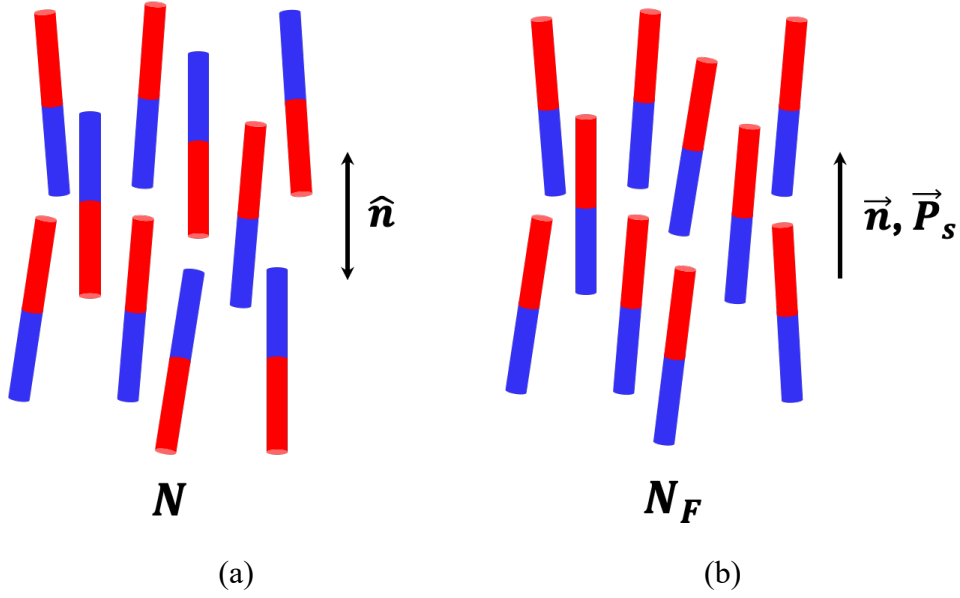

*Figure 2: Schematic illustration of nematic (N) and ferroelectric nematic ($N_F$) liquid crystal phases. Red (blue) parts indicate positively (negatively) charged areas corresponding to molecular dipoles. There are equal number of dipoles pointing up and down in the N phase (a), while most molecular dipoles point upward (or downward) in the $N_F$ phase, leading to a spontaneous polarization $\vec{P}_s$ along the director $\vec{n}$ (b).*

Here we will show that the time dependence of the polarization current of FNLC films is very similar to that observed in ferroelectric crystals, confirming their transient NC behavior as well. The ease of the film deposition at ambient temperature and the possibility to stabilize them by polymer networks, make FNLCs attractive to test them in low-power field effect transistors (FET) or in flexible organic electrochemical transistor (OECT)[32] devices.

## II. Results and Discussion

A 3.8 $\mu m$ thick cell with two 8 $nm$ thin gold electrodes of active area 2 $mm$ × 2$mm$ is filled with a room temperature ferroelectric nematic liquid crystal mixture KPA-02[33] at 85 °C by capillary action to form a Metal-Ferroelectric-Metal (MFM) configuration as shown in the inset to Figure 3(a). KPA-02 is a mixture containing 60 wt.% of a newly synthesized ferroelectric nematic liquid crystal compound 4-[(2,6-difluoro-4-cyanophenoxy) carbonyl] phenyl 2-n-propoxybenzoate (RT12155), and 40 wt.% of a commercially available nematic liquid crystal preparation, HTG-135200-100 (clearing point $T_c$ = 97 °C) from HCCH. On cooling, KPA-02 has an isotropic to ferroelectric nematic (N$_F$) phase transition at 47 °C. The N$_F$ phase is stable below room temperature.

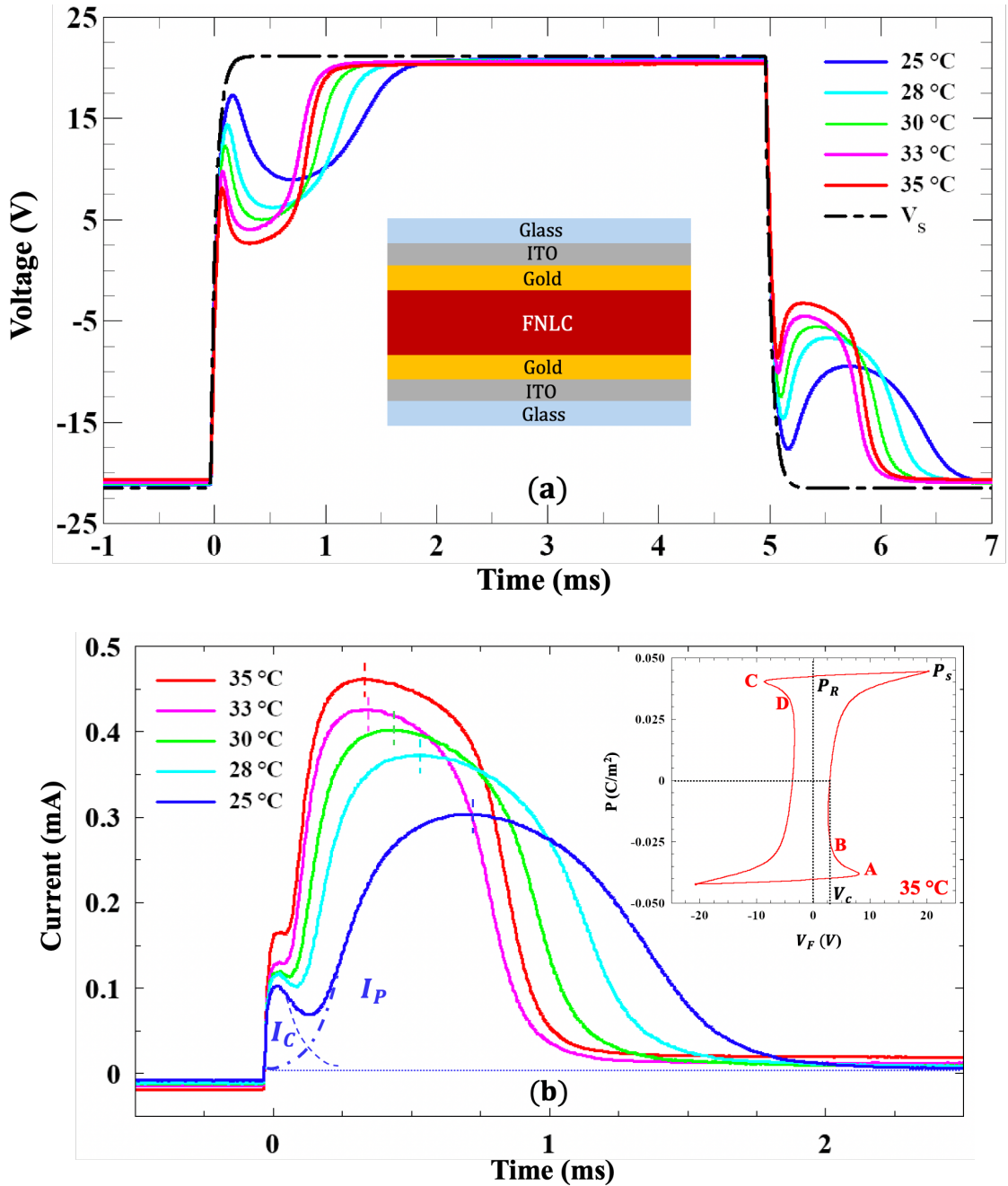

Figure 3: *Summary of electric measurements of a 2 mm × 2mm area, 3.8 μm thick KPA-02 cell. (a): Time dependence of the source voltage V$_S$ and the voltage V$_F$ dropping on the ferroelectric nematic MFM film illustrated in the inset at different temperatures for R= 40 kΩ resistance used in a circuit shown in Figure 2(c). (b): Time dependences of the electric current at various tempeeratures. Inset shows P(V$_s$) at 35 °C and indicating the coercive voltage V$_c$, the remnant polarization (P$_R$) and saturated polarization (P$_s$).*

The time dependences of the applied square wave voltage ($V_S$) applied to the FNLC film and the voltage $V_F$ measured in the FNLC film are shown in Figure 3(a). $V_F$ initially follows the input pulse but as the polarization is switching from negative to positive, it shows a negative peak in time interval increasing from $1\ ms$ at $35\ °C$ to $1.9\ ms$ at $25\ °C$. Similarly, when $V_S$ switches from positive to negative, $V_F$ shows positive peaks due to the polarization switching from positive to negative value. Inset in Figure 3(a) illustrates the side view of the FNLC sandwich cell. Figure 3(b) shows the time dependence of the current $I(t) = [V_S(t) - V_F(t)]/R$ flowing through the FNLC after switching from negative to positive sign of the applied voltage $V_s$. The charge per area, i.e., the ferroelectric polarization, can be calculated as $P(t) = \frac{\int I_P(t)dt}{A} = \frac{\int (V_S(t)-V_F(t))dt}{RA}$. The $P(t)$ curves show maxima marking the time when most director is switching. These switching times increase from $0.25\ ms$ at $35\ °C$ to $0.8\ ms$ at $25\ °C$. These values are an order of magnitude larger than measured in ferroelectric crystals under 20-times larger fields. [5] As the switching time in ferroelectric is inversely proportional to the applied field, we expect similar switching times under comparable fields in ferroelectric nematics and ferroelectric crystals.

The $P(V_F)$ function is plotted in the inset of Figure 3(b) at 35 °C. This (and all $P(V_F)$ curves measured at other temperatures) shows a hysteresis. For sections $AB$ and $CD$ the slope of the curve is negative, indicating that the capacitance is negative in these regions. The saturation value of spontaneous polarization ($P_s$) of KPA-02 at 35 °C is $P_s = 0.045\ C/m^2$, which is in agreement with previous measurements[33]. The remnant polarization ($P_R = 0.042$) measured at zero voltage is only slightly smaller than $P_s$, as usual for ferroelectric crystals. Note here the capacitive charge $Q_C = C \cdot V = \varepsilon_o \varepsilon_r \frac{A}{l} \cdot V$ is negligible compared to $Q_P = P \cdot A$. This means that $\varepsilon_r \ll \frac{P \cdot l}{\varepsilon_o V} \approx \frac{0.04 \cdot 4 \cdot 10^{-6}}{20 \cdot 10^{-11}} \approx 200$ in the frequency $f\left(\frac{1}{0.1} ms \approx 10 kHz\right)$ corresponding to the inverse rise time. The value of $Q_C$ could be best estimated at 25°C (see blue dashed curve) where $\varepsilon_r(\sim 10\ kHz) \approx 40$. From the ohmic leakage current $I_\Omega \approx 8.3\ \mu A$ measured at 25 °C under steady $V_s = 20\ V$, one can also calculate the electric conductivity $\sigma = \frac{I_\Omega \cdot l}{V_s \cdot A} \approx 4 \cdot 10^{-7}\ S/m$. Due to the fluid nature of the $N_F$ phase, this value is clearly much larger than of ferroelectric crystals, although in principle it can be reduced by several order of magnitudes, as $< 10^{-10} S/m$ values are typical for commercially available liquid crystals.

We also measured $V_F(t)$ as function of amplitude of the applied sqare wave voltage $V_s$. It was found that only capacitive and ohmic current occur when the applied square wave voltage is smaller than the coercive voltage. Similarly, no transient negative capacity was obeserved even for $V_s > V_c$ if either the sign inversion of $V_s$ was repeated very quickly (in less than 0.2 ms), i.e., when there was not enough time for the ferroelectric polarization to flip from one direction to the other, or $V_s$ was simply switched off to zero. These confirm that the transient negative capacitance exists only when the polarization flips between two opposite directions.

Finally, we note that a number of ferroelectric nematic liquid crystal materials cannot be switched with fields applied between the film insulating substrates if the depolarization field $E_d$ is larger than the electric breakdown field. In those cases, to mitigate the formation of the depolarization field, most often in-plane electric fields are provided by indium tin oxide (ITO) layers deposited in one substrate and separated by a gap. This way the polarization switches between positive and negative states compatible both with the surface alignment and of the applied field. Although these materials may not be useful for field-effect transistors, we have checked some of their $V_s - V_F$ characteristics (see Figure S1-S3 of the Supporting Information) and found them possessing NC characteristics similar to that of PZT ferroelectric crystals and our KPA-02 FNLC discussed by us above.

To summarize, we have shown that similar to recent seminal observations in certain solid ferroelectric crystals, fluid ferroelectric nematic liquid crystal films also show transient negative capacitance while switching their ferroelectric polarization from parallel to the substrates toward the electric field applied. Although this switching is different from the sidewise motion of domains with opposite directions of polarization of ferroelectric crystals, the observed switching time per unit field are similar in FNLCs and in ferroelectric crystals. Albeit current FNLCs have much larger electric conductivity than of ferroelectric crystals, they have several advantages over ferroelectric crystals. For example, they are lightweight, can be prepared and operated at ambient temperature, the direction of the polarization can be set by simple surface alignment, and require much less electric field for switching of the polarization than of solid ferroelectrics. These properties make FNLCs attractive to test them in in flexible organic electrochemical transistor (OECT) devices such as Ionic liquid crystal elastomer based OECTs have recently shown to have excellent transistor properties. [34]

## III. Acknowledgments

This work was financially supported by the US National Science Foundation grant DMR-2210083

# Supporting Information

# Transient Negative Capacitance in Ferroelectric Nematic Liquid Crystals

Netra Prasad Dhakal[1,2], Alex Adaka[1,2], Robert J. Twieg[4], Antal Jákli[1,2,3]*

[1]Advanced Materials and Liquid Crystal Institute, Kent State University, Kent, OH 44242, USA

[2]Materials Science Graduate Program, Kent State University, Kent, OH 44242, USA

[3]Department of Physics, Kent State University, Kent, OH 44242, USA

[4]Department of Chemistry and Biochemistry, Kent State University, Kent, OH 44242, USA

*: Corresponding author: ajakli@kent.edu

Transient negative capacitance of two dingle component ferroelectric nematic liquid crystal (FNLC) materials have been investigated in in-plane switching geometry. One of the materials is 4-[(4-nitrophenoxy)carbonyl)] phenyl 2,4-dimethoxybenzoate (RM734), which is one of the prototypical FNLC material [1]. On cooling from the isotropic liquid phase (I) it transitions to a conventional nematic (N) phase at 188 °C then to ferroelectric nematic ($N_F$) phase at 133 ° which crystallizes (Cr) at 84 °C. The chemical structure of RM734 is shown in the top inset of Figure S1.

The other FNLC compound is *4-[(2,6-difluoro-4-cyanophenoxy) carbonyl] phenyl 2-n-propoxybenzoate* (RT12155), which is the main component of the room temperature KPA-02 FNLC mixture we reported in the main document. On cooling from the isotropic liquid phase RT12155 transitions directly to the $N_F$ phase at 79 °C. The material did not crystallize while cooling to room temperature but on long standing at room temperature it turned glassy. The chemical synthesis and its physical characterization are described in the Supporting Information of Reference [2]. The chemical structure of RT12155 is shown in the top inset of Figure S3.

As shown in the bottom inset of Figure S1, the in-plane switching cells have two glass substrates: one is a bare glass, the other one is coated with a 50 $nm$ thick indium tin oxide (ITO) layer. The ITO-coated glass substrate is patterned with two parallel electrodes with a separation of 1 $mm$. The glass substrates are cleaned, dried, and spin-coated with polyimide PI2555 (HD MicroSystems) and baked. The PI2555 layers on both the glass substrates are then rubbed uniformly with the Rayon YA-19-R rubbing cloth (Yoshikawa Chemical Company, Ltd, Japan) with a pressure of 500 $N/m^2$. The rubbing gives uniform director alignment in the N and $N_F$ phases along the buffing direction. The two substrates are assembled with 10 μm spacers. The liquid crystal is filled inside the assembled cell by capillary filling method in the isotropic phase and cooled to the $N_F$ phase at the rate of 1°C/min. A rectangular voltage pulse ($V_S$) of frequency 100 Hz is generated from the Agilent 33120A function generator which is then multiplied by the KEPKO multiplier (BOP 500M). The input voltage pulse ($V_S$) is measured from channel 1 of the oscilloscope (Keysight 3024A). The liquid crystal (LC) cell is connected with a series resistance (R). The output voltage ($V_F$) across the liquid crystal cell is measured using channel 2 of the same oscilloscope. We use identical cables for both channels and make the cable length as short as possible to reduce the parasitic effect.

Time dependences of the a rectangular source voltage ($V_S$) and the ferroelectric voltage ($V_F$), measured using a circuit shown in Figure 1c of the main document of RM734 at 120 °C with R=200 kΩ resistance is seen in the main pane of Figure S1. When $V_S$ changes sign quickly, $V_F$ first switches from negative to positive, then decreases within 1 $ms$, and then increases up to about 2 $ms$ and gets saturated. This behavior is typical for ferroelectric crystals with transient negative capacitance. It is also similar to that we have found for KPA 02 cells switching under voltages applied between ITO coated substrates and shown in Figure 3a of the main document. The upper inset of Figure S1 shows the molecular structure of RM734, while the lower inset represents the schematics of the in-plane switching liquid crystal cell with green arrows showing the rubbing direction.

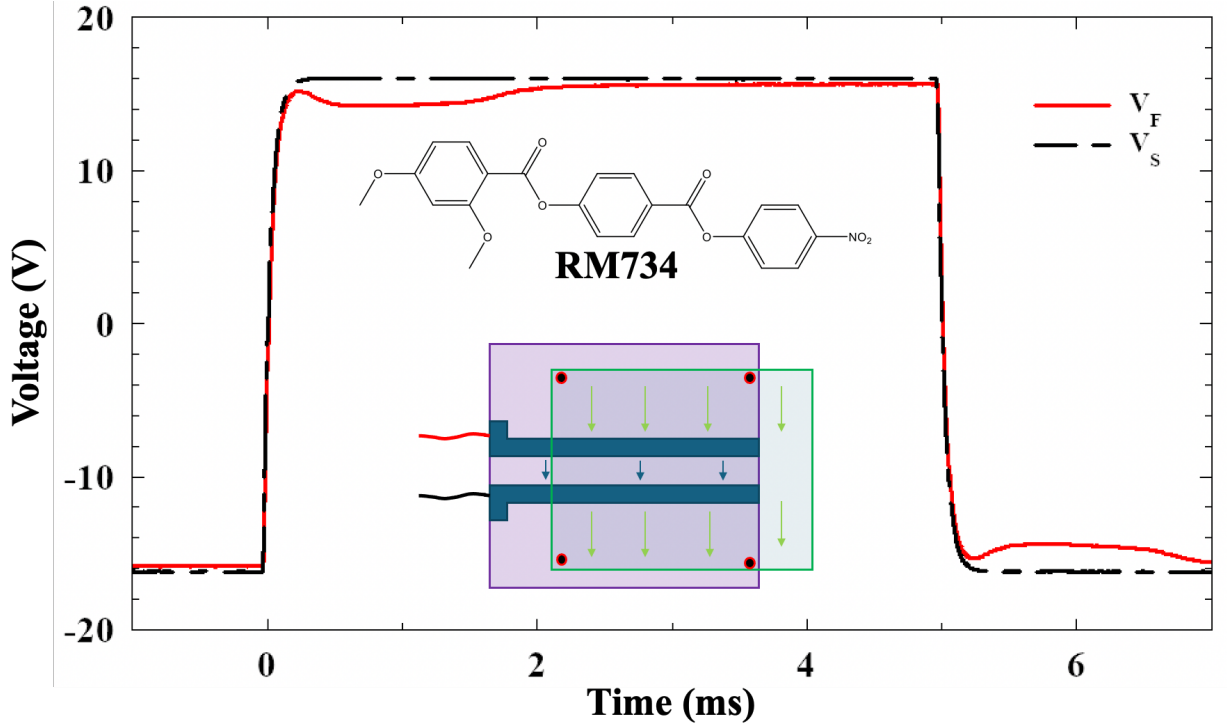

*Figure S1: Summary of the results obtained on in-plane switching10 μm RM734 film at 120 °C. Main pane: time dependences of the a rectangular source voltage ($V_S$) and the ferroelectric voltage ($V_F$), measured using a circuit shown in Figure 1c of the main document with R=200 kΩ resistance. Upper inset: Molecular structure of RM734. Lower inset: Schematics of the in-plane switching liquid crystal cell. Green arrows show the rubbing direction.*

Figure S2 represents the time dependence of the electric current flowing through RM734 at 120 °C. It is similar to that found in KPA-02 in sandwich cell geometry (see Figure 3b). The main difference is that the relative contribution of the capacitive current is slightly higher and the contribution of the ohmic current is much higher due to the higher electric current related to the high (120 °C) temperature. The saturated polarization calculated from the polarization current peak centered at around $1\ ms$ is about $P = 0.06\ C/m^2$, in agreement with previous results [3]. As seen in the inset to Figure S3, the voltage dependence of $P$ shows a hysteresis with negative slopes proving negative capacitance similar to that we found for KPA-02 in Figure 3b.

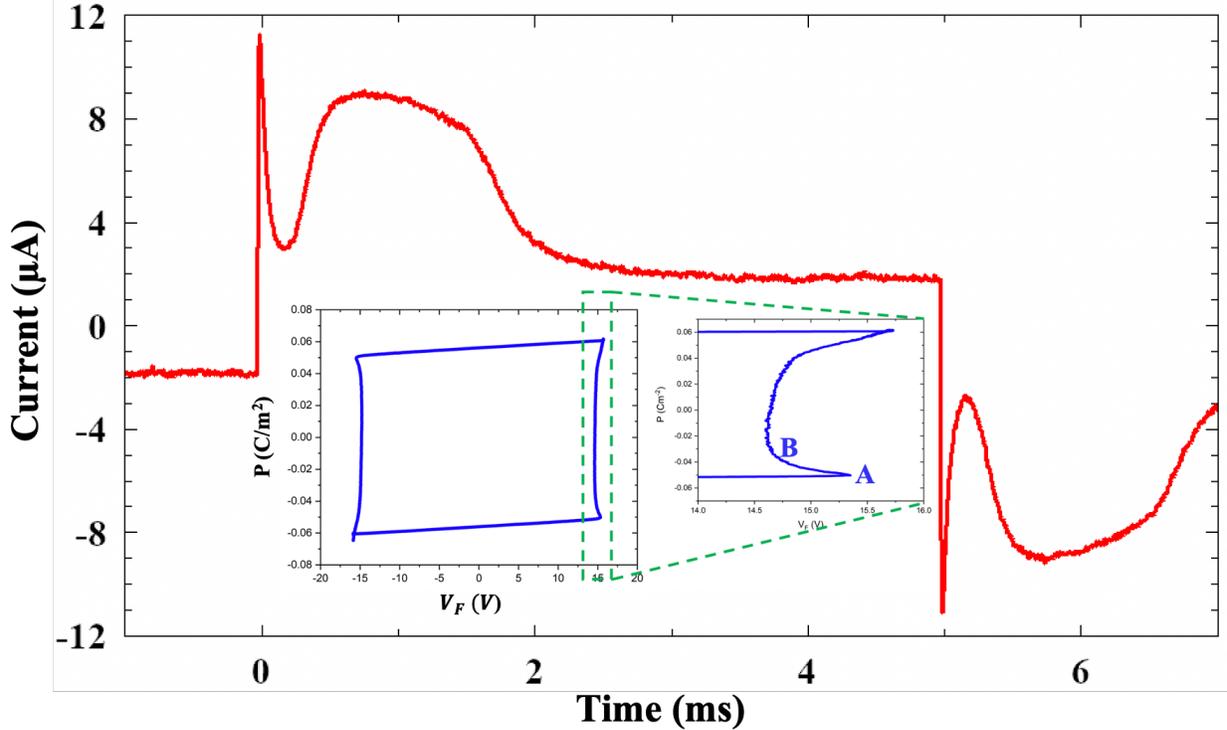

*Figure S2: Time dependence of the electric current flowing through the ferroelectric nematic material RM734 at 120 °C. Inset shows the voltage dependence of P with enlarged edge showing the negative slopes between points A and B.*

Another cell of thickness 4.8 $\mu m$ is filled with RT12155 liquid crystal and cooled to 70 °C. The time dependence of the rectangular applied voltage ($V_S$) pulse and the measured voltage drop ($V_F$) on the ferroelectric nematic cell are plotted in Figure S3. The higher input voltage is required due to the high viscosity of the material. The upper inset shows the molecular structure of RT 12155, while the lower inset shows the time dependence of the current flowing through the RT12155 cell. Both $V_F(t)$ and $P(t)$ are similar to what we obsrved in KPA-02 sandwich cell (see Figure 3b) and on RM734 in-plane switching cell (see Figure S2), indicating again negative capacitance.

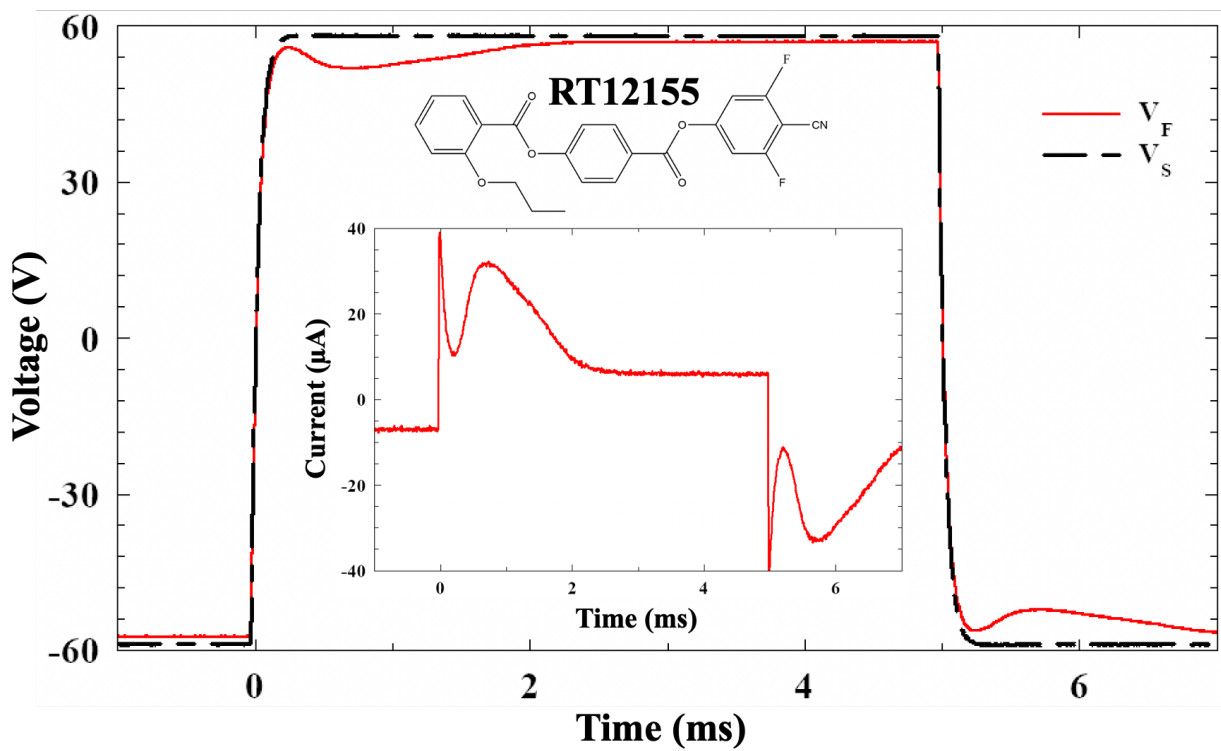

*Figure S3: Time dependences of the a rectangular source voltage ($V_S$) and the measured ferroelectric voltage ($V_F$ a 4.8 µm thin RT12155 in-plane switching cell at 70 °C for R= 200 kΩ. Upper inset: molecular structure of RT12155. Lower inset: Time dependence of the electric current flowing through the ferroelectric nematic material.*